\documentclass{optica-article}

\journal{opticajournal} % for journals or Optica Open

\articletype{Research Article}
\usepackage{cleveref}

\usepackage{lineno}
\usepackage{xr}
\begin{document}

\title{Enhancing Fluorescence Lifetime Parameter Estimation Accuracy with Differential Transformer Based Deep Learning Model Incorporating Pixelwise Instrument Response Function}

\author{Ismail Erbas,\authormark{1,2,*} Vikas Pandey,\authormark{1,2}  Navid Ibtehaj Nizam,\authormark{1,2} Nanxue Yuan,\authormark{1,2} Amit Verma,\authormark{3} Margarida Barosso,\authormark{3} Xavier Intes,\authormark{1,2}}

\address{\authormark{1}Department of Biomedical Engineering, Rensselaer Polytechnic Institute, Troy, NY 12180, USA\\
\authormark{2}Center for Modeling, Simulation, and Imaging in Medicine, Rensselaer Polytechnic Institute, Troy, NY 12180, USA\\
\authormark{3}Department of Molecular and Cellular Physiology, Albany Medical College, Albany, NY 12208, USA}

\email{\authormark{*}erbasi@rpi.edu} %% email address is required; see note below about the corresponding author designation

% use {asbstract*} to suppress the copyright line. Copyright information will be added in production

\begin{abstract*} 
Fluorescence Lifetime Imaging (FLI) is a critical molecular imaging modality that provides unique information about the tissue microenvironment, which is invaluable for biomedical applications. FLI operates by acquiring and analyzing photon time-of-arrival histograms to extract quantitative parameters associated with temporal fluorescence decay. These histograms are influenced by the intrinsic properties of the fluorophore, instrument parameters, time-of-flight  distributions associated with pixel-wise variations in the topographic and optical characteristics of the sample. Recent advancements in Deep Learning (DL) have enabled improved fluorescence lifetime parameter estimation. However, existing models are primarily designed for planar surface samples, limiting their applicability in translational scenarios involving complex surface profiles, such as \textit{in-vivo} whole-animal or imaged guided surgical applications. To address this limitation, we present MFliNet (Macroscopic FLI Network), a novel DL architecture that integrates the Instrument Response Function (IRF) as an additional input alongside experimental photon time-of-arrival histograms. Leveraging the capabilities of a Differential Transformer encoder-decoder architecture, MFliNet effectively focuses on critical input features, such as variations in photon time-of-arrival distributions. We evaluate MFliNet using rigorously designed tissue-mimicking phantoms and preclinical \textit{in-vivo} cancer xenograft models. Our results demonstrate the model's robustness and suitability for complex macroscopic FLI applications, offering new opportunities for advanced biomedical imaging in diverse and challenging settings.

\end{abstract*}

%%%%%%%%%%%%%%%%%%%%%%%%%%  body  %%%%%%%%%%%%%%%%%%%%%%%%%%
\section{Introduction}
Fluorescence lifetime imaging (FLI) is a powerful molecular imaging technique with high sensitivity and the ability to provide unique signatures with high specificity \cite{becker2012fluorescence,verma2024fluorescence}. Fluorescence lifetime and its associated parameters enable multiplexing studies \cite{ochoa2022computational,rudkouskaya2020multiplexed,kumar2018tomographic} and can report on numerous unique biological signatures, including micro-environmental parameters, protein conformations, metabolic states, protein-protein interactions, and/or ligand-target engagement  \cite{wang2017rapid,suhling2015fluorescence,yuan2024antibody,verma2024using}. FLI has known constant growth over the last three decades, with a significant acceleration in its dissemination thanks to the availability of a user-friendly FLI microscope \cite{datta2020fluorescence,dmitriev2021luminescence,sherry2023near}.   In parallel, over the last two decades, FLI has found an increased utility in translational applications, ranging from the mesoscopic (mFLI)  \cite{gao2022design} to the macroscopic regime (MFLI) \cite{venugopal2010development,kumar2020macroscopic,berezin2010fluorescence}. Compared to microscopic implementations, mFLI and MFLI are significantly more challenging due to the requirement of using Near Infrared (NIR) fluorophores for deeper tissue penetration. As fluorophores are red shifted, it is typical that their lifetimes are shorter (nanosecond (ns) or sub-nanosecond compared to few nanoseconds in the visible) \cite{berezin2010fluorescence} whereas large-format detectors exhibit low quantum efficiency (a few percent only) \cite{chavez2023characterization,nothdurft2012fluorescence}. Hence, quantifying lifetime and its associated parameters can be challenging due to very short fluorescence decays and/or low photon counts \cite{rudkouskaya2020quantification, rudkouskaya2018quantitative,marcu2012fluorescence,yuan2024experimental}. Unlike microscopic imaging, where the sample preparation allows for precise control over the imaging plane, mFLI, and MFLI samples can exhibit a large depth of field (DOF). These lead to significant variations in the time of arrival of the acquired data, which needs also be taken into account for accurate lifetime quantification. This is especially important in clinical systems, such as endoscopic or fluorescence-guided surgical instruments in which the tissue profiles can lead to DOF variations of a few centimeters.

To address these challenges, understanding the underlying methodology for estimating lifetime parameters becomes important. In mFLI and MFLI, lifetime parameters can be estimated by deconvolving the temporal point spread function (TPSF) and instrument response function (IRF). TPSF is the temporal histogram of the acquired fluorescence photons exiting the surface of the sample after a pulse excitation. The IRF represents the temporal response of the imaging system to pulsed illumination \cite{chen2019vitro}. Considering the complexity involved in estimating FLI parameters across diverse imaging conditions, fast and advanced data processing techniques are necessary to enhance both the precision and efficiency of these analyses. Recently, the field has seen a shift toward rapid, fit-free deep learning (DL) methodologies to alleviate the computational burden and reliance on user expertise, typically associated with methods such as nonlinear least squares fitting (NLSF) for FLI parameter estimation \cite{pandey2024deep,Nizam2024}. This advancement makes real-time FLI a possibility \cite{erbas2024compressing,erbas2024unlocking}, driven by the development of novel DL methods that eliminate traditional time-consuming computational approaches. FLI-Net, a DL model developed for FLI parameter estimation \cite{smith2019fast}, is used to analyze FLI data quickly, producing 2D quantitative images of the lifetimes and corresponding parameters directly without requiring manual parameter adjustments, outputting 2D quantitative images of the lifetime parameters directly. FLI-Net is versatile in terms of the imaging domain, including visible and near-infrared (NIR) imaging, making it adaptable to a large range of biomedical applications. FLI-Net takes the TPSF as an input and outputs FLI parameters. The experimental IRF was used in data generation of the training data; hence, it was represented in the TPSFs. However, FLI-Net was not designed to analyze pixel-wise IRFs while it predicts the lifetime parameters.

Despite the advancements in deep learning for FLI data analysis, the lack of pixel-wise IRF considerations poses limitations. The IRF integrates both the excitation part (including the laser source temporal profile) and detection (including the electronic limited reaction time) aspects of the optical setup. The complex broadening or distortion of the intrinsic fluorescence decay is caused by the detection part of the IRF from imaging system characteristics; however, the temporal offset in the IRF is caused by the photon time-of-arrival delays caused by the distance between the imager and sample surface \cite{yuan2024experimental}. Hence, topographic variations in the sample surface can lead to variations in delays in photon arrival times per pixel. In such cases, it is important to incorporate pixel-wise IRF in the FLI data processing pipeline.  To address these limitations and consider the essential role of the IRF in accurate FLI parameter estimation, we propose a novel deep-learning approach tailored specifically for processing FLI data. 

Following the developments in DL models, we leverage herein the ability of transformers to handle sequential data. Generally, transformers have a natural ability to capture long-range connections within data \cite{wen2022transformers,khan2022transformers,zaheer2020big,vaswani2017attention}. In the context of FLI, they can effectively identify and learn the relationships between the TPSF and IRF for accurate lifetime estimation. Furthermore, the self-attention mechanism in transformers allows them to focus on the most relevant parts of the input data for making predictions \cite{vaswani2017attention}. Recently, Differential Transformer (DIFF Transformer) as been proposed as a new approach to amplify attention to the relevant task while canceling noise \cite{ye2024differentialtransformer}. MFliNet architecture is based on a novel decoder layer design that integrates the DIFF Transformer for the first time in literature, improving the model's adaptability to account for shifts caused by variations in the DOF. The DIFF Transformer employs a differential attention mechanism, which calculates attention scores as the difference between two separate softmax attention maps, effectively canceling noise. This cancellation mechanism encourages sparse attention patterns, intensifying the focus on relevant contextual data while reducing distractions caused by irrelevant input. Analogous to noise-canceling systems, this approach mitigates the problem of over-allocating attention to non-critical information, a common issue with traditional transformers. DIFF Transformers outperform conventional transformers across various domains, especially in long-context modeling, key information retrieval, and robustness against variability in input structure. In the context of Time-Resolved FLI, these characteristics enable the detection of critical patterns even in low signal-to-noise scenarios, achieving higher accuracy and robustness compared to standard transformer architectures. These advancements mark a significant progression in FLI methodologies, offering a more effective tool for handling complex biological and imaging variability.

\section{Methods}
\subsection{Imaging setup}
All experimental data used in this work were captured on our MFLI system, detailed information about which can be found in \cite{venugopal2011small}. Briefly, the system uses a large-format Intensified Charge-Coupled Device (ICCD) camera (Picostar HR, LaVision GmbH, Germany), for wide-field detection over a $8 \times6$ cm$^2$  in combination with a Digital micro-mirrors device (DMD) (DLi 4110, Texas Instruments, TX, USA), for wide-field illumination. As an excitation source, we used a tunable Ti-Sapphire laser (Mai Tai HP, Spectra-Physics, CA, USA), which delivers $100$ femtosecond pulses at $80$ MHz. A gate width of 300 picoseconds (ps) and gate delay of $40$ ps were used for capturing time-resolved fluorescence decays (for \textit{in-vivo} and \textit{in-vitro} experiments) with a total of 176 time points, which is referred to as a number of gates (G=176). An emission filter at $740 \times 10$ nm (FF01-740/13-25, Semrock, IL, Rochester, NY, USA) is used to capture the TPSFs, with the laser set at a $700$ nm wavelength.  

\subsection{Generation of training data and classical Fluorescence lifetime processing}
Fluorescence Lifetime decay follows exponential decay. Depending on the number of components present in the sample, the decay kinetics can be described by a combination of multi-exponential functions. Most FLI imaging experiments involve up to two components, hence a bi-exponential model is typically used. The two-component or bi-exponential model also includes mono-exponential cases (where fractional amplitudes $A_R$ are one or zero). Mathematically, the TPSF is the convolution of the IRF and the fluorescence decay associated with the lifetime parameters as shown in Eq. \ref{eq:conv}, where lifetime decays are denoted as $\tau_1$, $\tau_2$, and $A_R$ is the amplitude fraction.
\begin{equation}
TPSF(t) = IRF(t) \ast \left( A_Re^{-\frac{t}{\tau_1}} + (1-A_R)e^{-\frac{t}{\tau_2}} \right)
\label{eq:conv}
\end{equation}
The \textit{in-silico} data used for training and validating the proposed model was generated using Eq. \ref{eq:conv}. Initially, time-resolved fluorescence lifetime images with dimensions of $28\times28$ pixels were generated by using the MNIST dataset. Fluorescence decays were generated for a range of lifetime values commonly used in NIR applications: $0.2$ ns to $0.8$ ns for $\tau_1$ (short-lifetime component) and $0.8$ ns to $1.5 $ ns for $\tau_2$ (long-lifetime component). The range of the $A_R$ (fraction amplitude) was set from 0\% to 100\%, respectively (both bounds corresponding to mono-exponential cases). To ensure that our simulated data accurately represents experimental applications, pixel-wise IRFs were used. To capture experimental IRFs, a white diffused paper was placed on the imaging table and illuminated using the DMD with an excitation wavelength of 700 nm. The reflected light was captured using a neutral density (ND) filter. Subsequently, each TPSF was generated by convolving randomly selected IRF from the dataset with simulated fluorescence decay profiles. To approximate the noise characteristics of real-world measurements, system-derived noise, including read-out noise, dark noise, etc., as explained in \cite{venugopal2011small}, was incorporated into the simulated TPSFs. This approach ensures that the simulated data closely matches the noise dynamics observed in the actual system.

To evaluate and compare the model's performance in the absence of experimental ground truth, we used the NLSF method, which is commonly used to estimate the FLI parameters described in Eq. \ref{eq:conv}.  Traditionally, FLI parameters are estimated from experimental data through iterative fitting optimization methods such as the NLSF, which incorporates the Levenberg-Marquardt algorithm \cite{lakowicz2006principles}, or center of mass (CMM) analysis \cite{li2011video}. For our NLSF analysis, we utilized a software named AlliGator \cite{chen2017alligator}, allowing adjustments and constraints on fitting parameters, including short and long lifetimes, fraction amplitudes, and offsets, depending on experimental conditions. We selected between single and double exponential decay models according to the complexity of our datasets. AlliGator also provides an option for offset correction when there is a mismatch between the TPSF and IRF. We evaluated the importance of offset correction in our NLSF analysis by comparing data with and without this feature in our phantom experiments. Additionally, we benchmarked our results against those obtained using FLI-Net and  to provide a comprehensive evaluation of our approach in the context of established methodologies. Furthermore, to assess the impact of the DIFF Transformer model, we compared our results with those from a transformer model of same architecture and parameters, trained on the same dataset.

\subsection{Deep learning network architecture}
\label{sec:deep-learning-architecture}

\begin{figure}[ht!]
\centering\includegraphics[width=0.8\linewidth]{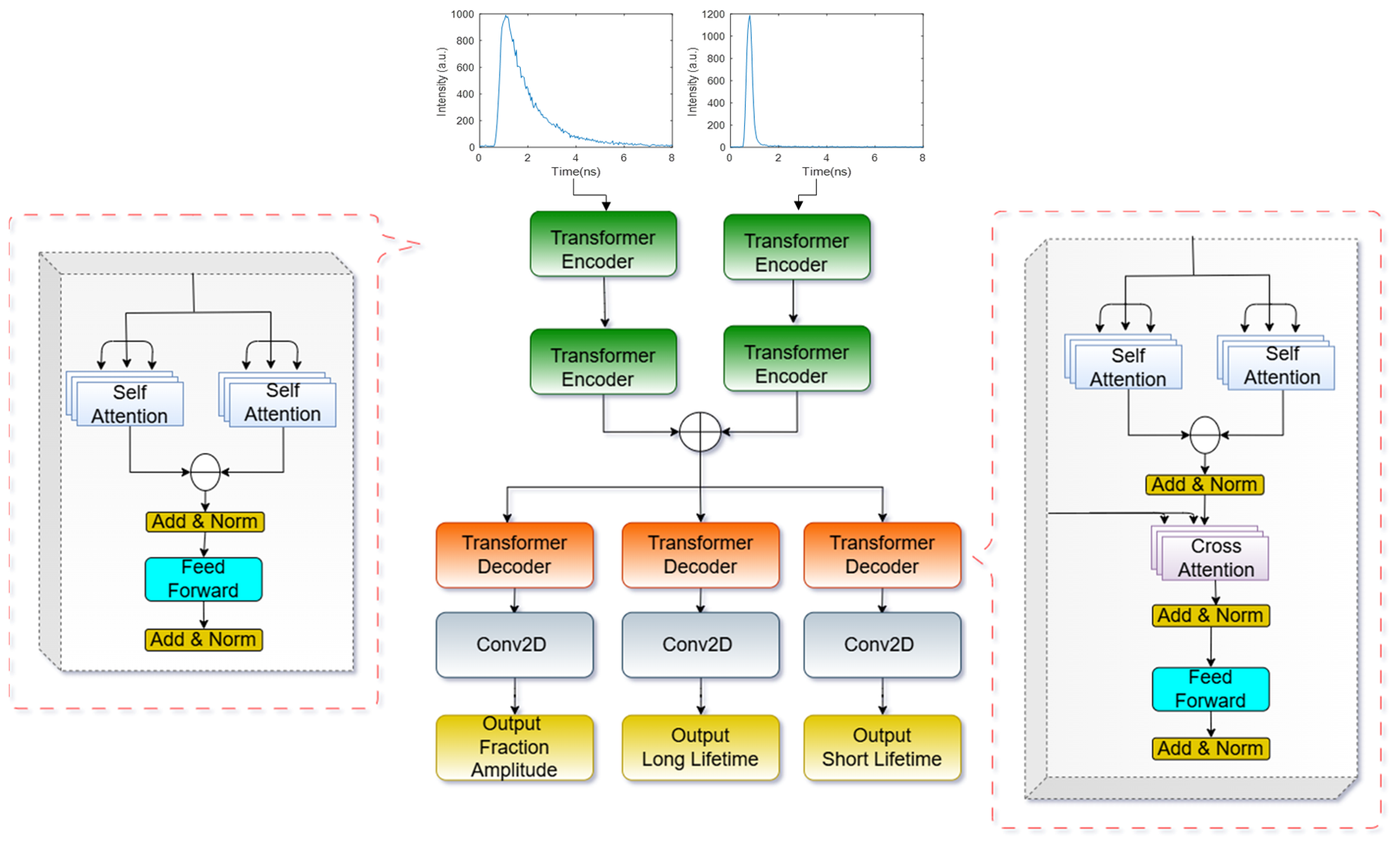}
\caption{Proposed transformer-based deep learning network architecture}
\label{fig:model}
\end{figure}

MFliNet is a novel architecture designed for FLI parameter estimation, particularly effective under varying IRFs. The model leverages a DIFF Transformer framework, incorporating a unique differential attention mechanism to enhance feature extraction while mitigating the effects of noise. The theoretical background on the differential attention mechanism is detailed in \cite{ye2024differentialtransformer}.

At the core of MFliNet is the differential attention mechanism, which computes attention scores by contrasting two separate multi-head attention outputs. Specifically, the mechanism calculates the difference between two attention maps, scaled by a parameter $\lambda$, to focus on relevant patterns and suppress noise. The differential attention is defined as:

\begin{equation}
    \text{DiffAttn}(X) = \left( \text{softmax}\left( \frac{Q_1 K_1^\top}{\sqrt{d_k}} \right) V_1 - \lambda \cdot \text{softmax}\left( \frac{Q_2 K_2^\top}{\sqrt{d_k}} \right) V_2 \right),
\end{equation}

where:

\begin{itemize}
    \item $Q_1, K_1, V_1$ are the query, key, and value matrices for the first attention head, derived from the input $X$ using learned projections.
    \item $Q_2, K_2, V_2$ are the query, key, and value matrices for the second attention head, also derived from $X$.
    \item $d_k$ is the dimensionality of the key vectors.
    \item $\lambda$ is a learnable scalar parameter that balances the contributions of the two attention maps.
\end{itemize}

The queries, keys, and values are computed as:

\begin{equation}
    Q_i = X W_i^Q, \quad K_i = X W_i^K, \quad V_i = X W_i^V, \quad \text{for } i = 1, 2,
\end{equation}

where $W_i^Q$, $W_i^K$, and $W_i^V$ are the learned weight matrices for the $i$-th attention head.

This differential attention mechanism enhances the model's ability to focus on critical features by emphasizing the differences between two attention outputs, effectively reducing the impact of noise. The parameter $\lambda$ controls the degree to which the second attention output is subtracted from the first.

Each input sequence passes through two stacked encoder blocks, each comprising a differential attention layer, followed by layer normalization and a feed-forward network (FFN) with SwiGLU activation. The encoder block operates as follows:

\begin{align}
    \text{Attention Output} &= \text{DiffAttn}(X), \\
    \text{Add \& Norm}_1 &= \text{LayerNorm}(X + \text{Attention Output}), \\
    \text{FFN Output} &= \text{FFN}(\text{Add \& Norm}_1), \\
    \text{Encoder Output} &= \text{LayerNorm}(\text{Add \& Norm}_1 + \text{FFN Output}),
\end{align}

where the feed-forward network is defined as:

\begin{equation}
    \text{FFN}(x) = \left( \text{Swish}(x W_1 + b_1) \odot (x W_2 + b_2) \right) W_3 + b_3,
\end{equation}

with $W_1, W_2, W_3$ being learned weight matrices, $b_1, b_2, b_3$ bias vectors, $\odot$ representing element-wise multiplication, and $\text{Swish}$ being the activation function defined as $\text{Swish}(x) = x \cdot \sigma(x)$, where $\sigma(x)$ is the sigmoid function.

The decoder blocks incorporate both self-attention and cross-attention mechanisms. The self-attention layer within the decoder uses the differential attention mechanism to capture intra-sequence relationships. The cross-attention layer aligns the decoder's inputs with the encoder outputs, integrating information from both inputs. The decoder block operates as:

\begin{align}
    \text{Self-Attn Output} &= \text{DiffAttn}(X), \\
    \text{Add \& Norm}_1 &= \text{LayerNorm}(X + \text{Self-Attn Output}), \\
    \text{Cross-Attn Output} &= \text{Attention}(\text{Add \& Norm}_1, E, E), \\
    \text{Add \& Norm}_2 &= \text{LayerNorm}(\text{Add \& Norm}_1 + \text{Cross-Attn Output}), \\
    \text{FFN Output} &= \text{FFN}(\text{Add \& Norm}_2), \\
    \text{Decoder Output} &= \text{LayerNorm}(\text{Add \& Norm}_2 + \text{FFN Output}),
\end{align}

where $E$ represents the encoder outputs from the corresponding input sequence, and the standard attention mechanism is defined as:

\begin{equation}
    \text{Attention}(Q, K, V) = \text{softmax}\left( \frac{Q K^\top}{\sqrt{d_k}} \right) V.
\end{equation}

MFliNet's architecture includes three parallel output pathways, each dedicated to predicting one of the FLI parameters: short lifetime, long lifetime, and fractional amplitude. Each pathway processes the decoder outputs through additional layers to refine the predictions.

The final outputs are obtained by reshaping the decoder outputs and applying a convolutional layer with kernel size $1 \times 1$, using the Exponential Linear Unit (ELU) activation function and L2 regularization to prevent overfitting:

\begin{equation}
    \text{Output}_i = \text{Conv2D}_{\text{ELU}}(\text{Reshape}(\text{Decoder Output}_i)), \quad \text{for } i = 1, 2, 3,
\end{equation}

where $\text{Conv2D}_{\text{ELU}}$ denotes a 2D convolutional layer with ELU activation and L2 regularization applied to the reshaped decoder output corresponding to each parameter.

Training was conducted using the Adam optimizer with an adaptive learning rate starting from 0.001. The loss function for each output branch was Mean Squared Error (MSE). The dataset comprised 2,000 samples (totaling 1,568,000 generated time-resolved photon signals and corresponding IRFs), with 10\% reserved for validation. The model's design allows it to capture both local and global patterns in the data, effectively modeling variations in time-domain signals and encoding the relationships between inputs and FLI parameters.

\subsection{Phantom preparation}
\begin{figure}[ht!]
\centering\includegraphics[width=\linewidth]{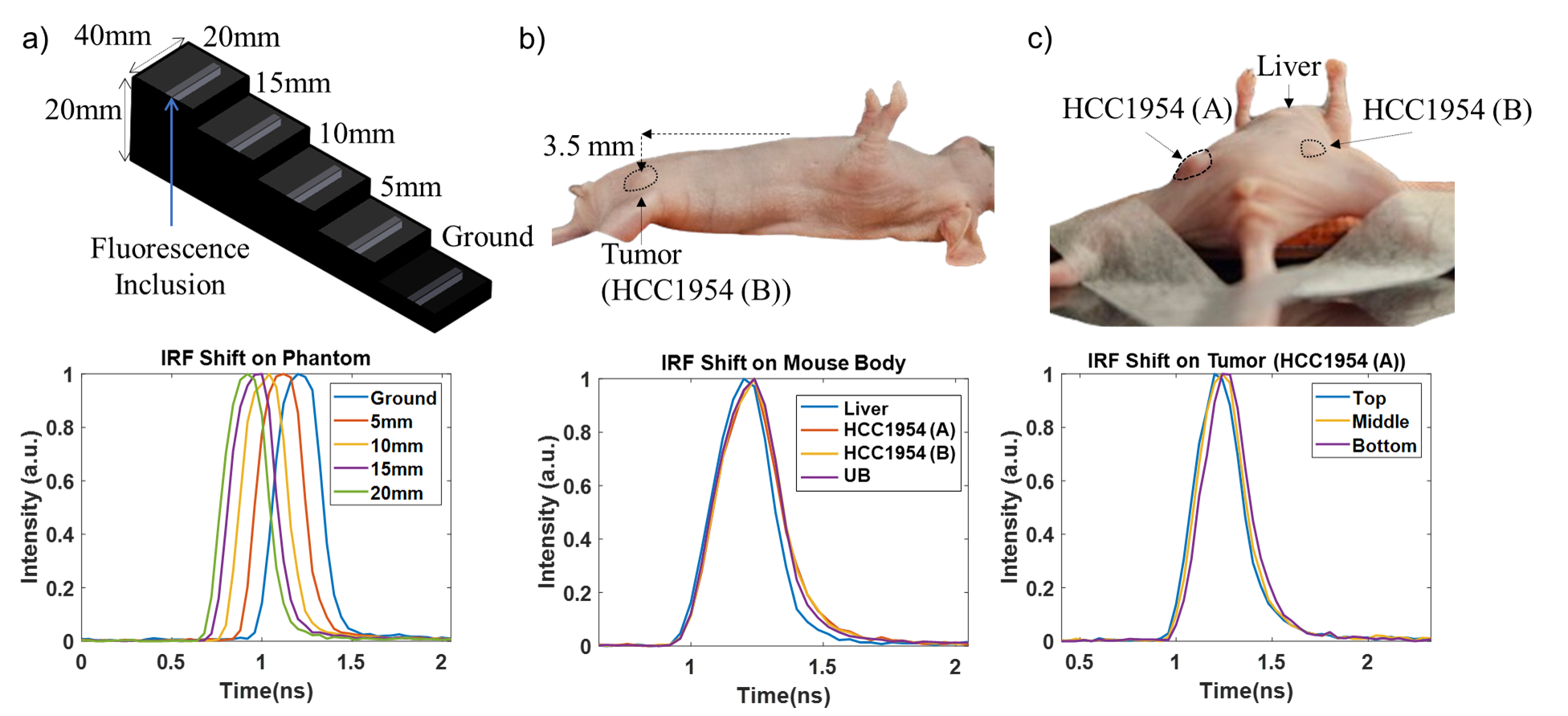}
\caption{Illustration of designed 3D step ladder phantom and the IRF shifts as a result of variations in height: (a) display of the 3D phantom (up) and plots of average of IRFs from each height (down); (b) a side view of a mouse, highlighting height differences between anatomical regions (up) and IRF plots of randomly selected pixels on liver, urinary bladder (UB), tumors (down);  (c) A distal ventral view of the mouse highlighting the tumors and the liver (up) and the IRF plots of the randomly selected pixels on the left tumor (down). }
\label{fig:irfshift}
\end{figure}

For experimental validation, we designed a step ladder phantom to introduce variations in sample-detector distance, as depicted in Figure \ref{fig:irfshift}(a). A 3D printable case was designed to accommodate five discrete containers arranged at various heights: ground level, $5$ mm, $10$ mm, $15$ mm, and $20$ mm.  Each container was crafted with dimensions of $40\times40 \times 10$ mm$^{3}$ to accommodate tissue-mimicking phantoms. The phantoms were made with agar constituting 1\% of the total volume ($80$ cm$^{3}$). To prepare the phantoms, agar was first fully dissolved in distilled water by heating, and then allowed to cool slightly before further processing. The optical properties of the phantoms (absorption coefficient ($\mu_a$) of $0.005$ mm$^{-1}$ and reduced scattering coefficient ($\mu_s'$) of $1$ mm$^{-1}$) were controlled through the addition of India Ink and intralipid solutions, to provide absorption and scattering contrasts respectively \cite{chavez2024multimodal}. In each ladder step, a specific area was designated for the placement of a cuboidal fluorescence embedding, with dimensions of  $5 \times 5 \times 40$ mm$^{3}$. This embedding consisted of Alexa Fluor 700 dye dissolved in phosphate-buffered saline to achieve a concentration of $20$ $\mu$M. The embeddings were placed at a depth of $1$ mm from the surface of each phantom.

To examine the offset variation across different heights on the phantom and in live intact animals, we plotted the pixelwise IRFs for comparison as shown in  Figure \ref{fig:irfshift}. For each specified height in the step ladder phantom, we plotted the average IRFs in Figure \ref{fig:irfshift} (a).  Moreover, we also illustrated the IRFs of various anatomical regions in live intact animals, including the liver, urinary bladder (UB), and tumors, in Figure \ref{fig:irfshift}(b).  Lastly, in  Figure \ref{fig:irfshift}(c), we examined the variability of the IRF within a single tumor. This plot contains IRFs on three points within a tumor: the top, middle, and bottom. The top refers to an IRF from the upper region of the tumor, the middle corresponds to an IRF from the central area in terms of height, and the bottom shows an IRF from the lowest part of the tumor. 
\subsection{\textit{In-vivo} experiment}
\label{sec:invivo}
For \textit{in-vivo} MFLI imaging experiments, we imaged HER2+ breast tumor xenografts HCC1954 in athymic nude mice. The cell line was sourced from ATCC (Manassas, VA, USA) and maintained in RPMI 1640 media enriched with 10\% fetal bovine serum (ATCC) and $50$ units/mL/ $50$ µg/mL penicillin/streptomycin from ThermoFisher Scientific (Waltham, MA, USA). We initiated tumor xenografts by subcutaneously injecting $5\times10^6$ HCC1954 cells suspended in PBS and mixed in a $1:1$ ratio with Cultrex BME (R\&D Systems Inc, Minneapolis, MN, USA) into the inguinal mammary fat pads of female athymic nude mice aged 4 weeks (CrTac: NCR-Foxn1nu, Taconic Biosciences, Rensselaer, NY, USA). Tumors were monitored daily for 4 weeks. The mouse was administered with a retro-orbital injection of AF700 conjugated with Meditope Trastuzumab (MDT-TZM) (MDT-TZM-AF700) at $20$ µg and AF750 conjugated with MDT-TZM (MDT-TZM-AF750) at $40$ µg in a $2:1$ acceptor to donor ratio through staggered injection \cite{verma2024using}. Donor injection was performed 2 hours ahead of acceptor injection through the retro-orbital route. MFLI Imaging was conducted 24 hours post-injection. Throughout the imaging process, the mouse was anesthetized with isoflurane, and the body temperature was maintained with a Rodent Warmer X2 (Stoelting, IL, USA). All animal procedures were conducted with the approval of the Institutional Animal Care and Use Committee (IACUC) at both Rensselaer Polytechnic Institute and Albany Medical College. The animal facilities of both institutions have been accredited by the American Association for Accreditation for Laboratory Animals Care International. 

\section{Results}
\begin{figure}[ht!]
\centering
\includegraphics[width=\linewidth]{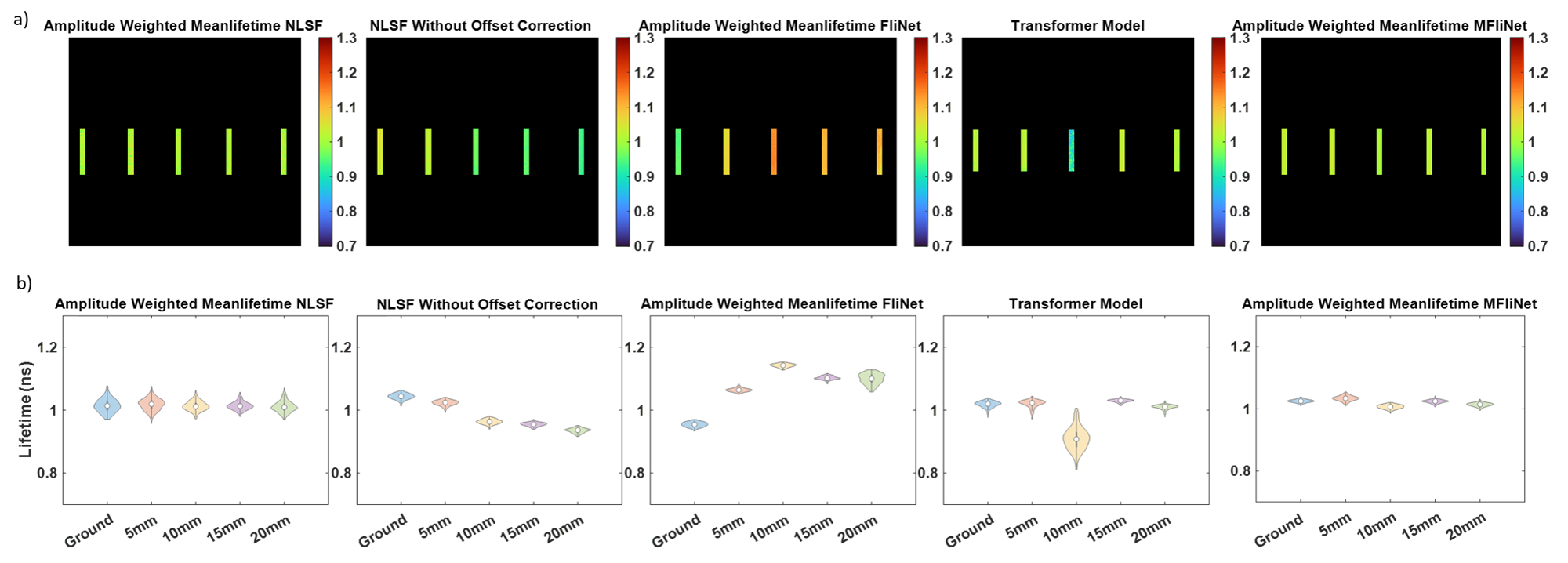}
\caption{Phantom experiment results. a) Image overlay of the lifetime estimation results, b) Violin plots of NLSF analysis, FLI-Net, transformer model and MFliNet}
\label{fig:phantom}
\end{figure}

We conducted the ladder phantom experiment designed to validate the MFliNet model under controlled conditions that mimic biological tissues. Figure \ref{fig:phantom} shows the phantom experiment results where the analysis was done using four methods: NLSF, FLI-Net, transformer model and MFliNet. To compare the precision and stability of each method under varying conditions reflective of real-world applications, results were evaluated across five different heights: ground, $5$ mm, $10$ mm, $15$ mm, and $20$ mm. A $160$ ps shift in the IRF was observed from ground level to a height of $20$ mm, with a shift of approximately $40$ ps for each $5$ mm increment in height. For simplicity in comparison, the amplitude-weighted average lifetime was calculated using Eq. \ref{eq:meanLifetime}, for all outputs.
\begin{equation}
\tau_M =  \left( A_R\tau_1 + (1-A_R)\tau_2 \right)
\label{eq:meanLifetime}
\end{equation}
NLSF analysis using pixel-wise IRF showed consistency in lifetime estimation across all tested heights. The mean fluorescence lifetime values obtained by NLSF were clustered around $1.01 \pm 0.02$ ns. NLSF without offset correction results deteriorated with each increase in height. At the ground level, NLSF without offset correction began with a mean value of $1.04 \pm 0.01$ ns. However, as the height increased, a steady decline in lifetime estimation was observed, reaching a mean value of $0.93 \pm 0.01$ ns at $20$ mm.  FLI-Net, in contrast, demonstrated a wider variation in estimated fluorescence lifetime values. At the ground level, it reported a mean value of $0.96 \pm 0.01$ ns, which was lower than the NLSF values. As the distance increased, FLI-Net's estimations deviated further, peaking at $1.14 \pm 0.01$ ns at $10$ mm and estimating $20$ mm with a mean value of $1.10 \pm 0.02$ ns. The transformer model showed variable performance across different heights; at 10 mm, a decrease was observed with a mean value of $0.91 \pm 0.04$ ns, while estimations at other heights were closer to the NLSF values. In contrast, MFliNet, showed closer results with NLSF, where the mean values were within the same range as NLSF. Moreover, in terms of processing speed, NLSF took approximately 6 hours to analyze 598 pixels (covering only the tumor area), whereas MFliNet analyzed the entire dataset of 90,480 pixels in just 63 seconds.

\begin{figure}[ht!]
\centering\includegraphics[width=12cm]{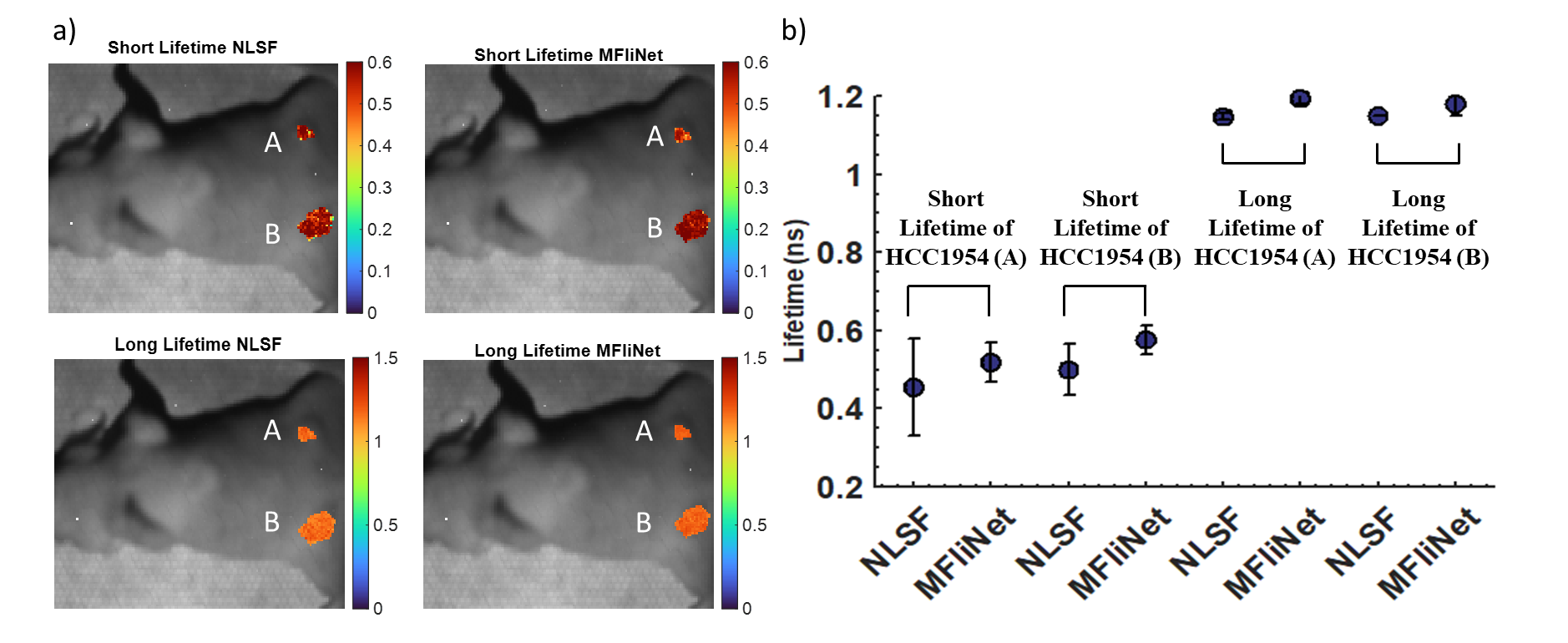}
\caption{Comparison of \textit{in-vivo} results for both NLSF and MFliNet a) Image overlays of the short and long-lifetime results for both NLSF and MFliNet b) plot of means and standard deviations of the predicted lifetime values of both methods}
\label{fig:invivo}
\end{figure}
 
Following the phantom studies, \textit{in-vivo} experiments were conducted using the HER2+ breast tumor xenograft model in mice to evaluate the model's performance in a more complex, biologically variable environment. The experimental results, illustrated in Figure \ref{fig:invivo}, demonstrate the comparative analysis of NLSF and MFliNet in HER2+ HCC1954 tumor xenograft mice model. For the smaller tumor (HCC1954 (A)), the NLSF method reported a short fluorescence lifetime of $0.56 \pm 0.06$ ns and a long lifetime of $1.17 \pm 0.03$ ns. The MFliNet showed a comparable short lifetime of $0.52 \pm 0.05$ ns and a long lifetime of $1.19 \pm 0.02$ ns. In the case of the larger tumor (HCC1954 (B)), the NLSF method reported a short fluorescence lifetime of $ 0.55 \pm 0.07$ ns and MFliNet showed a comparable short lifetime of $0.58 \pm 0.04$ ns. For the long lifetime, both methods again yielded closely aligned values: $1.16 \pm 0.04$ ns for NLSF and $1.18 \pm 0.03$ ns for MFliNet. 

\section{Discussion and Conclusion}

In this study, we introduced MFliNet, a novel deep learning model based on the DIFF Transformer architecture, to address the challenges of accurate FLI parameter estimation, particularly in complex and variable biological environments. Our results, as illustrated in Figure \ref{fig:irfshift},  demonstrate shifts in IRFs at varying heights, highlighting how each organ's unique geometry and composition contribute to IRF offsets. This variation in the IRF offset underscores the challenge of accurately estimating the FLI parameters and the necessity for MFliNet, which can adapt to these complexities. The integration of pixel-wise IRF analysis within MFliNet specifically addresses the effects of surface irregularities on early photon arrival times, which is often overlooked in other DL models. By comparing MFliNet with a standard transformer model of same architecture and parameters, we demonstrated that the differential attention mechanism inherent in the DIFF Transformer enhances the model's ability to focus on relevant features while suppressing noise, leading to improved FLI parameter estimation. 

Comparative analysis indicates that MFliNet not only matches the accuracy of NLSF analysis but also enhances processing speed. MFliNet eliminates the need for manual user dependency and extensive user training, making it better suited for real-time applications. In addition, as shown in the phantom experiment, an increasing trend in lifetime estimations from the FLI-Net suggests a distance-related bias, which reflects an underlying limitation in the model's ability to account for variations in time-of-flight. The effect of the IRF offset on lifetime estimation is further validated through NLSF analysis without using the offset correction, where lifetime estimations result in noticeable declines, potentially leading to systematic underestimations of fluorescence lifetimes and inaccuracies in diagnostics.

The significance of these improvements is particularly relevant in complex imaging environments such as fluorescence-guided surgery (FGS),  where the understanding of these variables can significantly impact the quality of imaging and, consequently, the surgical outcomes. Potential integration of MFliNet with existing FGS systems can lead to the development of advanced surgical guidance systems that offer real-time, precise imaging for cancer surgery \cite{ochoa2023assessment}. Moreover, the capabilities of MFliNet extend beyond clinical applications, offering potential benefits in various research applications. In drug development, for instance, MFliNet's enhanced accuracy could be used to determine drug-target interactions more precisely, thus accelerating the development of therapeutics by providing clearer insights into molecular engagements. Additionally, in biological research, the improved measurement accuracy of molecular interactions facilitated by MFliNet could foster a deeper understanding of cellular functions and disease mechanisms. This could open new avenues for exploring and developing targeted therapies.  This work contributes to the field by providing a robust and efficient tool for FLI parameter estimation, with potential applications in clinical diagnostics, fluorescence-guided surgery, and various biomedical research areas.

\begin{backmatter}
\bmsection{Funding}
This work was supported by the National Institutes of
267 Health (NIH) Grants R01CA237267, R01CA250636, R01CA271371, \& R01CA250636-02S1.

\bmsection{Acknowledgment}
The authors thank Dr. Xavier Michalet for his support with the AlliGator software.

\bmsection{Disclosures}
The authors declare no conflicts of interest.
\medskip
\bmsection{Data availability} Data underlying the results presented in this paper are not publicly available at this time but may be obtained from the authors upon reasonable request.

\bigskip

\end{backmatter}

\label{sec:refs}

%%%%%%%%%%%%%%%%%%%%%%% References %%%%%%%%%%%%%%%%%%%%%%%%%

%%%%%%%%%% If using BibTeX:
\bibliography{sample}

%%%%%%%%%% If preparing manually:
% \begin{thebibliography}{1}
% \newcommand{\enquote}[1]{``#1''}

% \bibitem{Zhang:14}
% Y.~Zhang, S.~Qiao, L.~Sun, Q.~W. Shi, W.~Huang, L.~Li, and Z.~Yang,
%   \enquote{Photoinduced active terahertz metamaterials with nanostructured
%   vanadium dioxide film deposited by sol-gel method,}
%   {\protect\JournalTitle{Optics Express}} \textbf{22}, 11070--11078 (2014).

% \bibitem{Optica}
% {Optica}, \enquote{{Optica Publishing Group},}
%   \url{http://www.opg.optica.org}.

% \bibitem{FORSTER2007}
% P.~Forster, V.~Ramaswamy, P.~Artaxo, T.~Bernsten, R.~Betts, D.~Fahey,
%   J.~Haywood, J.~Lean, D.~Lowe, G.~Myhre, J.~Nganga, R.~Prinn, G.~Raga,
%   M.~Schulz, and R.~V. Dorland, \enquote{Changes in atmospheric consituents and
%   in radiative forcing,} in \enquote{Climate Change 2007: The Physical Science
%   Basis. Contribution of Working Group 1 to the Fourth Assesment Report of
%   Intergovernmental Panel on Climate Change,}  S.~Solomon, D.~Qin, M.~Manning,
%   Z.~Chen, M.~Marquis, K.~B. Averyt, M.~Tignor, and H.~L. Miler, eds.
%   (Cambridge University Press, 2007).

% \end{thebibliography}

\end{document}